\documentclass[amsmath,comment,amssymb,twocolumn,showpacs,prb]{revtex4}
\usepackage{graphicx}
\usepackage{amssymb}
\usepackage{epsfig}
\usepackage{dcolumn}
\usepackage{bm}

\newcommand{\be}{\begin{equation}}
\newcommand{\ee}{\end{equation}}
\newcommand{\bea}{\begin{eqnarray}}
\newcommand{\eea}{\end{eqnarray}}

\newcommand{\nn}{\nonumber }

\begin{document}
\title{Mean field theory of superglasses}
\author{Xiaoquan Yu}
\affiliation{International School for Advanced Studies (SISSA), via Bonomea 265, 34136 Trieste, Italy}
\author{Markus M\"uller}
\affiliation{The Abdus Salam International Center for Theoretical Physics, Strada Costiera 11, 34151 Trieste, Italy}

\date{\today }

\begin{abstract}
We study the interplay of superfluidity and glassy ordering of hard core
bosons with random, frustrating interactions. This is motivated by bosonic systems such as amorphous supersolid,
disordered superconductors with preformed pairs and helium in porous media. We analyze the fully
connected mean field version of this problem, which exhibits three low
temperature phases, separated by two continuous phase transitions: an
insulating, glassy phase with an amorphous frozen density pattern, a
non-glassy superfluid phase and an intermediate phase, in which both types
of order coexist.
We elucidate the nature of the phase transitions, highlighting in
particular the role of glassy correlations across the superfluid-insulator
transition. The latter suppress superfluidity down to $T=0$, due to the
depletion of the low energy density of states, unlike in the standard BCS
scenario.
Further, we investigate the properties of the coexistence (superglass)
phase. We find anticorrelations between the local order parameters and a
non-monotonous
superfluid order parameter as a function of $T$. The latter arises due to
the weakening of the glassy correlation gap with increasing temperature.
Implications of the mean field phenomenology for finite dimensional
bosonic glasses with frustrating Coulomb interactions are discussed.
\end{abstract}

\maketitle



\section{Introduction}

Bosons in random environments occur in a variety of experimentally relevant systems, ranging from cold atomic gases, superconductors and quantum liquids. The superfluidity of Helium-4 ($^4{\rm He}$) in porous media was one of the first phenomena observed in this type of systems,~\cite{Cao, Reppy92} featuring an interesting competition between Bose-Einstein condensation and localization by random potentials.~\cite{Yamamoto}

In more recent years, supersolidity in crystalline $^4{\rm He}$~\cite{KimChan1,KimChan2,Reppy06} has been reported. It soon became clear that defects of the crystalline order and amorphous solids sustain more robust supersolidity, spurring the idea that disorder, or even glassy order, may be a crucial element in understanding the superfluid part of those systems.~\cite{Supersolidityanddisorder,superglassphaseinHe4,SupersolidityinHe,Disorderinducedsuperfluidity,thesupersolidityofhelium}

A recent experiment~\cite{Hunt09} reported indeed that the supersolidity in $^4{\rm He}$ is accompanied by the onset of very slow glassy relaxation. This suggested that an amorphous glass with a superfluid component is forming, a state of matter which was dubbed a "superglass". These experimental results have motivated several theoretical investigations into the possibility and nature of such amorhously ordered and yet supersolid systems.~\cite{Superglass, Gingras, Zamponi, Leggettbound,Laura,Ying-Jer}

Similar questions as to the coexistence and interplay of glassy density ordering and superfluidity arises in disordered superconducting films, which feature disorder- or field-driven superconductor-to-insulator quantum phase transitions.~\cite{Goldmanreview}
In several experimental materials, this transition appears to be driven by phase fluctuations of the order parameter rather than by the depairing of electrons, suggesting that the transition can be described in terms of bosonic degrees of freedom only.~\cite{Hebard} This had lead to the dirty boson model~\cite{FisherBoseglass} and the notion of the Bose glass,~\cite{GiamarchiSchulz} in which disorder and interactions lead to the localization of the bosons, while the system remains compressible.\cite{SItransition, MaLee}
In this context, the term "glass" refers mostly to the amorphous nature of the state rather  than to the presence of slow relaxation and out-of equilibrium phenomena due to frustrated interactions. However, if superconductivity develops in a highly disordered environment, such frustration may add in the form of Coulomb interactions between the charged carriers, which may become important, as screening is not very effective. It is well known that in more insulating regimes, strong disorder and Coulomb interactions may induce a glassy state of electrons (the Coulomb glass).~\cite{DaviesLeeRice, PankovDobrosavljevic, MuellerIoffe} It is therefore an interesting question whether such glassy effects can persist within the superconducting state of disordered films. A memory dip in resistance versus gate voltage (similar to the conductance dip in insulators, which is considered a smoking gun of electron glassiness~\cite{Ben-Chorin93}) was reported as a possible indication of such a super-glassy state in Ref.~\onlinecite{Goldmaninversedip}, even though doubts about its intrinsic nature were raised later on.~\cite{GoldmanInversedip_revoke}

The recent developments in ultra cold atoms~\cite{Roati} open new ways to studying bosonic atoms in the presence of both interactions and disorder.
Those can exhibit superfluid or localized, and potentially also glassy phases, especially if the interactions are sufficiently long ranged and frustrated, as  is possibly the case for dipolar interactions.

Motivated by these experimental systems, we study a solvable model of bosons subject to disorder and frustrating interactions, as proposed previously in Ref.~\onlinecite{Gingras}. This solvable case provides insight into the possibility of coexistence of superfluidity and glassy density order, as well as into the nature of the coexistence phase (the superglass). In particular, for the considered mean field model we prove the existence of a superglass phase. This complements the numerical evidence for such phases provided by quantum Monte Carlo investigations in finite dimensions~\cite{Gingras} and on random graphs.~\cite{Zamponi} Those were, however, limited to finite temperature, and could thus not elucidate the structure of the phases at $T=0$.
In contrast, the analytical approach allows us to understand the quantum phase transition between glassy superfluid and insulator, and the non-trivial role played by glassy correlations. Indeed it is an interesting feature of the considered model that despite of its infinite connectivity, it features an insulating, Anderson localized phase at sufficiently small but finite hopping strength. This is in contrast to non-glassy models where the localization and insulating behavior is lost in the large connectivity limit, unless the hopping is down-scaled logarithmically with the connectivity.~\cite{Abou-Chacra}

The model studied here is an interesting type of a quantum glass.~\cite{P Young} Like other canonical quantum glass models of mean field type, such as the random exchange Heisenberg model,~\cite{BrayMoore1980} the Ising spin glass in transverse field,~\cite{Goldschmidt,Cha96} and frustrated rotor models,~\cite{Ye93,Read95,Georges00} the glass phase of the superglass model breaks ergodicity, and will exhibit a large number of metastable states and associated slow relaxation and out-of-equilibrium dynamics. It has recently been pointed out that such mean field glasses can faithfully be realized by atoms in laser cavities, which are coupled at long distances through interactions via discrete photon modes.~\cite{Goldbart,StrackSachdev}

In the above mentioned canonical models the non-glassy phase is usually quantum disordered, having no broken symmetry. However, a new situation arises in the model studied here: The glassy ordering competes with a different order: the superfluidity, or the transverse (XY) ferromagnetism in a magnetic analogon. The microscopic coexistence of these two types of order, which we demonstrate to exist in our model, is rather non-trivial, as one might instead expect a first order transition between phases with either of the two orders.
At zero temperature, the quantum glass transition is induced by frustrated interactions, which win over weak quantum fluctuations.
While the quantum phase transition of such models is understood relatively well,~\cite{Sachdevbook} the deep glass phase with its collective gapless modes has not been much explored.~\cite{MuellerIoffe07, Andreanov}

Let us finally remark that the glassy, amorphous supersolid, which the "superglass" phase constitutes, is quite different from the type of supersolid proposed theoretically in the early seventies.~\cite{coexitingsuperfluidandbrokentranslationalsymmerty} In those scenario the bosons organize spontaneously on a lattice, which breaks translational symmetry, but is incommensurate with the boson density, allowing for vacancies to move through the solid. Our model considers instead bosons on a predefined lattice, on which an inhomogeneous density pattern establishes in the glass phase.

The remainder of the paper is organized as follows: In Sec.II we introduce the mean field model of a superglass.  We obtain the effective action for a single spin with both the replica method and the cavity approach in Sec.III, and introduce the key concept of the distribution of local fields. The self-consistency equations of the mean field theory are solved under static approximation, which is argued to be exact in parts of the phase diagram.
Sec.IV determines the instabilities towards forming glassy and superfluid order, and establishes the phase diagram featuring two quantum phase transitions at $T=0$. We also discuss the robustness of the phase diagram to random potential disorder. In Sec.V, we study the bulk of the "superglass" phase in more detail. We show that the glass and  superfluid order parameters are locally anti-correlated. Moreover, we find the superfluid order parameter to have an interesting non-monotonous behavior as a function of $T$.
The implications of this mean field analysis for realistic finite dimensional models, e.g. with frustrating long range Coulomb interactions, will be discussed  in Sec.VI. Some detailed derivations are collected in the appendix.

\section {Model}
We consider the fully connected model of hard-core bosons with random pairwise interactions between all bosons,
\bea
\label{OH}
\ H=-\sum_{i<j}V_{ij} n_{i}n_{j} +\sum_i\epsilon_i n_i -\frac{t_{b}}{N} \sum_{i<j}\left(b^{\dag}_{i}b_{j}+b^{\dag}_{j}b_{i}\right).
\eea
Here, $n_{i}=b^{\dag}_{i} b_{i}$ is the number operator on site $i$, and the
hard core constraint limits $n_i$ to assume values $0$ or $1$. $b_{i}(b^{\dag}_{i})$ denote the annihilation (creation) operators for a hard-core boson at site $i$. $V_{ij}$ is a quenched disorder with Gaussian distribution of zero mean and variance $V^2/N$, $\epsilon_i$ describes a quenched disorder potential for the bosons, and $t_{b}/N$ is the unfrustrated hopping strength between any pair of sites. The scaling of the couplings with $N$ is chosen so as to yield a non-trivial thermodynamic limit for $N \to \infty$.

In the absence of hopping, the model becomes classical and is equivalent to the Sherrington-Kirkpatrick spin glass (SK) model~\cite{SK} in a random field. It is well-known that except for lowering the transition temperature, the random fields do not alter the low temperature properties of the glass phase. We thus restrict our attention mostly to a slightly simpler model proposed in Ref.~\onlinecite{Gingras}, which corresponds to a special choice of the $\epsilon_i$:
\bea
\label{boseHam}
H&=&-\sum_{i<j}V_{ij}(n_{i}-1/2)(n_{j}-1/2)\nn\\
&&-\frac{t_{b}}{N}\sum_{i<j}\left(b^{\dag}_{i}b_{j}+b^{\dag}_{j}b_{i}\right).
\eea
Similar fermionic mean field models have been studied in Refs.~\onlinecite{Pastor99,Dalidovich02}.
The identification $2n_{i}-1=s^{z}_{i} \in \{\pm 1 \}$, $ b^{\dag}_{i}=s^{+}_{i}, b_{i}=s^{-}_{i}$, allows us to map this model into a fully connected spin glass model with quantum fluctuations arising from non-random spin flip terms,
\bea
\label{Hamiltonian1}
\ H=-\sum_{i<j}J_{ij}s^{z}_{i}s^{z}_{j}-\frac{t}{N}\sum_{i<j}\left(s^{x}_{i}s^{x}_{j}+s^{y}_{i}s^{y}_{j}\right),
\eea
with the simple dictionary
\bea
\label{dict}
J_{ij} = \frac{V_{ij}}{4},\quad t= \frac{t_b}{2}.
\eea

For $t=0$, this Hamiltonian reduces to the SK model, which possesses a spin glass phase at low temperature, $T<T_{g}=J$. Without the Ising interactions, $J_{ij}=0$, the Hamiltonian turns into the mean field XY model, which has a superfluid (or XY ferromagnetic) phase at low temperatures ($T<T_{s}=t$).
In this paper we establish the phase diagram and study the properties of the bulk phases resulting from the competition of random density-density interactions and boson hopping (bosonic language) or equivalently random Ising interactions and ferromagnetic transverse coupling (spin language).

\section {Free energy and self-consistent equations}
\subsection {General formalism}

The disorder average of the free energy of the model (\ref{Hamiltonian1}) can be obtained using the replica method~\cite{MezardParisi}
\bea
\left<\log Z\right>_{J}=\lim_{n\rightarrow0}\frac{\left<Z^{n}\right>_{J}-1}{n},
\eea
where $Z$ is the partition function and $\left<...\right>_{J}$ indicates an average over the couplings $J_{ij}$.

Following a method introduced by Bray and Moore~\cite{BrayMoore1980} it is useful to represent the partition function as an imaginary time path integral:
\bea
Z^{n}&=&{\rm Tr} {\cal T}\exp \Biggl \{ \beta \int^{1}_{0}d\tau\sum_{a=1}^n\sum_{i<j} \Biggl[J_{ij}s^{z}_{ia}(\tau)s^{z}_{ja}(\tau) \nn\\
&& +\frac{t}{N}\Biggr(s^{x}_{ia}(\tau)s^{x}_{ja}(\tau)+s^{y}_{ia}(\tau)s^{y}_{ja}(\tau)\Biggr) \Biggr] \Biggr\},
\eea
where ${\cal T}$ orders the operators in decreasing order of their argument $\tau\in [0,1]$.
This "time" argument of  $s(\tau)$ merely serves us to define the time-ordering, while $s(\tau)$ denotes always the same Pauli matrix, independently of time.

Averaging over disorder and decoupling the spins on different sites using a Hubbard-Stratonovich transformation with the order parameter fields $Q_{ab}, M^x_a, M^y_a$, we obtain:

\bea
\label{PF}
\left<Z^{n}\right>_{J}&\propto&\int \prod_{a} dQ_{aa}(\tau,\tau^{\prime})\,dM^{x}_{a}(\tau)\,dM^{y}_{a}(\tau) \nn\\
&&\times \prod_{a<b} dQ_{ab}(\tau,\tau^{\prime})\exp\left(-N{\cal F}\right)
\eea
with
\bea
\label{replicafreeenergy}
{\cal F}&=&\frac{J^{2}\beta^{2}}{4}\int^{1}_{0}\int^{1}_{0}d\tau d\tau^{\prime}\nn\\
&&\times\left[\sum_{a\neq b} Q^{2}_{ab}(\tau,\tau^{\prime})+\sum_{a} Q^{2}_{aa}(\tau,\tau^{\prime})\right] \nn\\
&&+\frac{t\beta}{2}\int^{1}_{0}d\tau\sum_{a}\Biggl[ M^{x}_{a}(\tau)^{2}+M^{y}_{a}(\tau)^{2}\Biggr]\nn\\
&&-\log{\cal Z},
\eea
\bea
{\cal Z}={\rm Tr}{\cal T}\exp\left(-S_{\rm eff}\right),
\eea
\bea
S_{\rm eff}&=&{\bf -}\frac{J^{2}\beta^{2}}{2}\int^{1}_{0}\int^{1}_{0}d\tau d\tau^{\prime}\\
&&\times\Biggl[\sum_{a\neq b}Q_{ab}(\tau,\tau^{\prime})s^{z}_{a}(\tau)s^{z}_{b}(\tau^{\prime})\nn\\
&&+\sum_{a}Q_{aa}(\tau,\tau^{\prime})s^{z}_{a}(\tau)s^{z}_{a}(\tau^{\prime})\Biggr]\nn\\
&&-t\beta\sum_{a}\int^{1}_{0}d\tau\Biggl(M^{x}_{a}(\tau)s^{x}_{a}(\tau)+M^{y}_{a}(\tau)s^{y}_{a}(\tau)\Biggr). \nn
\eea

In the limit $N\rightarrow \infty$, the functional integral (\ref{PF}) is dominated by the saddle point of the replicated free energy ${\cal F}$, which satisfies
\bea
0&=&\frac{\delta {\cal F}}{\delta Q_{ab}(\tau,\tau^{\prime})} \Rightarrow
Q_{ab}(\tau,\tau^{\prime})=\left<{\cal T}s^{z}_{a}(\tau)s^{z}_{b}(\tau^{\prime})\right>_{\rm eff}\nn\\
&=&\left<s^{z}_{a}s^{z}_{b}\right>_{\rm eff} \equiv Q_{ab},
\\
0&=&\frac{\delta {\cal F}}{\delta Q_{aa}(\tau,\tau^{\prime})} \Rightarrow  Q_{aa}(\tau,\tau^{\prime})=\left<{\cal T}s^{z}_{a}(\tau)s^{z}_{a}(\tau^{\prime})\right>_{\rm eff}\nn\\
& \equiv & R(\tau,\tau^{\prime}),
\\
0&=&\frac{\delta {\cal F}}{\delta M^{x}_{a}(\tau)} \Rightarrow  M^{x}_{a}(\tau)=\left<s^{x}_{a}(\tau)\right>_{\rm eff}\equiv  M^x,
\\
0&=&\frac{\delta {\cal F}}{\delta M^{y}_{a}(\tau)} \Rightarrow  M^{y}_{a}(\tau)=\left<s^{y}_{a}(\tau)\right>_{\rm eff} \equiv M^y,
\eea
where $\left<...\right>_{\rm eff}$ denotes the average with respect to the effective action $S_{\rm eff}$ of a single site.
We have used that, as usual, the saddle point values of $Q_{a\neq b}$ and $M^{x,y}_a$ are independent of imaginary time, while $Q_{aa}(\tau,\tau')$ depends only on the imaginary time difference.~\cite{GrempelRozenberg} Furthermore, $Q_{aa}$ and $M^{x,y}_a$ do not depend on the replica index $a$. For $Q_{ab}$ we make the standard ultrametric ansatz, parametrized by a monotonous function $q(x)$ on the interval $x\in [0,1]$, which is well-known to describe successfully the SK model and other mean field glasses.\cite{MezardParisi}
We are free to choose coordinates in the $x,y$ plane such that the spontaneous magnetization $\vec{M}$ points in the $x$-direction, and thus we set $M^y=0$.

Note that $M^x\neq 0$ signals the presence of transverse (XY) order of the spins, that is, superfluidity of the hard core bosons, which breaks the $U(1)$ symmetry spontaneously. On the other hand, a non-constant value of $Q_{a\neq b}$ implies the spontaneous breaking of the replica symmetry, and thus the presence of a glass phase with many metastable states and non-trivially broken ergodicity. As long as we do not consider random field disorder, the breaking of replica symmetry coincides with the breaking of the Ising symmetry and is signalled by a nonzero value of $Q_{a\neq b}$.
We will see below that the $U(1)$ and the replica symmetries can be broken  simultaneously in a what has been called a "superglass phase" in Refs.~\onlinecite{Gingras, Zamponi}.

To find the location of a (continuous) glass transition, we expand the free energy to second order in $Q_{ab}$. We find an instability towards replica symmetry breaking, and thus  the emergence of a glassy density ordering of bosons, when
\bea
\beta J\int^{1}_{0} \int^{1}_{0} d\tau d\tau^{\prime}\left<{\cal T}s^{z}_{a}(\tau)s^{z}_{a}(\tau^{\prime})\right>_{\rm eff}&=&\beta J\int^{1}_{0}d\tau R(\tau)\nn\\
&=&1,
\eea
or
\bea
\label{instability of glass}
 J \chi^\parallel =1,
\eea
where $\chi^\parallel\equiv \chi_{zz}(\omega=0)$ is the zero-frequency limit of the longitudinal susceptibility. This condition is of course to be evaluated at $Q_{a\neq b}=0$.

On the other hand, a second order phase transition from the high temperature phase towards a superfluid state is indicated by the instability condition, which follows from $\partial^2{\cal F}/\partial M^2=0$:
\bea
\beta t\int^{1}_{0}\int^{1}_{0}d\tau d\tau^{\prime}\left<{\cal T} s^{x}_{a}(\tau)s^{x}_{a}(\tau^{\prime})\right>_{\rm eff}=1,
\eea
or
\bea
\label{instability of superfluidity}
 t \chi^\perp=1,
\eea
where $\chi^\perp \equiv \chi_{xx}(\omega=0)$ is the static transverse susceptibility.
These expressions must be calculated in the non-superfluid phase where $M=0$. In this regime the effective action $S_{\rm eff}$ is classical, which entails the further simplification $R(\tau)=1$. This feature is due to the suppression of quantum fluctuations  in the non-superfluid phase by factors of $1/N$, due to the scaling of the transverse coupling. It allows us to find the superfluid-insulator transition analytically, even at zero temperature, without solving a full quantum impurity problem. In particular, we immediately find that the transition from the disordered high temperature phase to a glassy phase is given by
\bea
T_g = J,
\eea
exactly as in the classical SK model. However, the glass transition line will be modified if it is preceded by a superfluid transition at higher temperature.

\subsection{Solution of the saddle point equations}
A full solution of the saddle point equations involves the solution of the problem of interacting replica as well as the evaluation of dynamical correlation functions with the effective action $S_{\rm eff}$, if $M\neq 0$ and the replica symmetry is broken as well.

Here we describe what steps an exact solution involves, and then discuss the approximations we will use to study parts of the phase diagram, especially the bulk of the superglass phase.

To describe a non-glassy superfluid phase, the replica structure is trivial, and one needs to solve the self-consistency equations
\bea
M=\left<s^x\right>_{\rm eff}, \quad
R(\tau)= \left<{\cal T} s^z(\tau) s^z(0)\right>_{\rm eff},
\eea
with effective action
\bea
S_{\rm eff}& =& -\frac{\beta^{2} J^2}{2} \int_0^1 \int_0^1 d\tau d\tau' s^z(\tau) R(\tau-\tau') s^z(\tau') \nn\\
&&-\beta tM \int_0^1  d\tau s^x(\tau).
\eea
These can be solved using techniques as used in dynamical mean field theory.~\cite{DMFT}

In a glassy phase the replica structure has to be taken into account. Assuming the standard ultrametric structure of the saddle-point matrix $Q_{ab}$, the above single-replica scheme has to be generalized to include a self-consistent distribution of frozen longitudinal fields $P(y)$ acting on a given replica. This captures the distribution of random frozen fields $y_i$ created by the exchange of sites $i$ with the frozen magnetization pattern with a spin glass state.~\cite{SD}
In practice this requires the simultaneous solution of
\bea
\label{SC}
m(y) &=&\left<s^z\right>_{S_{\rm eff}(y)}, \nn\\
m_{x}(y) &=&\left<s^x\right>_{S_{\rm eff}(y)}, \nn\\
M  &=& \int dy P(y)m_{x}(y), \nn\\
R(\tau)&= &\int dy P(y) \left<{\cal T}s^{z}(\tau) s^{z}(0)\right>_{S_{\rm eff}(y)},
\eea
where the effective single replica action in a frozen field $y$ reads
\bea
S_{\rm eff}(y)&=&-\frac{\beta^{2} J^2}{2} \int_0^1 \int_0^1 d\tau d\tau' \nn\\
&&\times s^z(\tau)\left[R(\tau-\tau')-q_{\rm EA}\right]s^z(\tau')\nn\\
&&-\beta tM \int_0^1 d\tau s^x(\tau)- \beta y \int_0^1  d\tau s^z(\tau).
\eea
The Edwards-Anderson order parameter $q_{\rm EA}$ characterizes the glassy freezing in a pure state of the glass and is given by
\bea
q_{\rm EA} =q(x=1)&=&  \int dy P(y)m^{2}(y).
\eea
As first derived by Sommers and Dupont,~\cite{SD} the frozen field distribution $P(y)\equiv P(y,x=1)$ is obtained from a self-consistent solution of the differential equations on the interval $x\in [0,1]$
\bea
\label{SD1}
\dot{m}(y,x)=-\frac{\dot{q}(x)}{2}\Biggl[m^{\prime\prime}(y,x)+2x m(y,x) m^{\prime}(y,x)\Biggr],\\
\label{SD2}
\dot{P}(y,x)=\frac{\dot{q}(x)}{2}\Biggl[P^{\prime\prime}(y,x)- 2x(m(y,x) P(y,x))^{\prime}\Biggr],
\eea
with
\bea
\label{IC}
m(y,x=1)= m(y), \quad
P(y,x=0)=\delta(y),
\eea
where dots and primes denote derivatives with respect to $x$ and $y$, respectively.
The solutions of these differential equations solve the saddle point equations of the replica free energy(\ref{replicafreeenergy}).~\cite{SD} The overlap function $q(x)$, which parametrizes the ultrametric matrix $Q_{ab}$ by the distance $x$ between replica, must obey the self-consistency relation
\bea
\label{SCE}
q(x)=\int P(y,x) m(y,x)^{2}.
\eea
Notice that Eqs.~(\ref{SD1}) and (\ref{SD2}) are the same as in a classical spin glass. The influence of quantum fluctuations enters through the boundary condition $m(y,x=1)\equiv m(y)$, where $m(y)$ was defined in Eq.~(\ref{SC}) These differential equations provide an elegant way of integrating out all spins except for one.~\cite{Duplantier}

Once the above scheme has been solved self-consistently, site-averaged observables such as the longitudinal magnetization are given by
\bea
M^z  \equiv \left< \frac{1}{N} \sum_is^z_i\right>&=&\int dy P(y)  \left<s^z\right>_{S_{\rm eff}(y)} \nn\\
&=& \int dy P(y) m(y).
\eea

The properties of the solution of these differential equations are well understood in several classical models exhibiting full replica symmetry breaking with continuous functions $q(x)$.~\cite{ParisiToulouse, CrisantiRizzo, Pankov, MuellerPankov}
The full solution of mean field quantum glasses in the ergodicity broken has not been analyzed in the literature so far. However, an analysis of the transverse field SK model shows that most features of the low temperature solution of $q(x)$ carry over rather naturally to the quantum case.~\cite{Andreanov} A salient new feature in the quantum case is the fact that full replica symmetry breaking implies marginal stability of the whole glass phase, which in turn ensures the presence of gapless collective excitations. The latter is very similar to what was found, e.g., in the threshold states of quantum p-spin models,~\cite{QTAP} or in the quantum dynamics of elastic manifolds, approximated with a replica symmetry breaking variational approach.~\cite{LeDoussalGiamarchi}

\subsection{Alternative derivation by a cavity approach}
The replica-diagonal part of the above scheme will become easier to understand, if we derive it in a cavity framework~\cite{replicawithoutreplicas} similarly to the derivation of the quantum analog of Thouless-Anderson-Palmer equations by Biroli and Cugliandolo.~\cite{QTAP} From a cumulant expansion in the couplings involving site $o$ it is easy to obtain the following effective action for the site $o$:
\bea
S_o^{\rm eff}&=&{\bf-}\frac{\beta^{2}}{2}\int^{1}_{0} \int^{1}_{0}d \tau d \tau^{\prime}\nn\\
&&\times s^{z}_{o}(\tau) \left[ \sum_{i}J_{oi}^{2} \left<s^{z}_{i}(\tau)s^{z}_{i}(\tau^{\prime})\right>^{o}_{c}\right]s^{z}_{o}(\tau^{\prime})\nn\\
&&-\beta \int^{1}_{0} d\tau \Biggl(h_o^{z}(\tau)s^{z}_{o}(\tau)+h^{x}(\tau)s^{x}_{o}(\tau)\Biggr).
\eea
Here
\bea
h_o^{z}(\tau)=\sum_{i}J_{oi}\left<s^{z}_{i}(\tau)\right>^{o}=\sum_{i}J_{oi}\left<s^{z}_{i}\right>^{o}
\eea
is the site-dependent longitudinal field, which does not depend on time, however. The index $o$ denotes a "cavity average", i.e. an average over the action of the system, in which the site $o$ has been removed. The subscript $c$ indicates a connected correlator. The effective transverse field,
\bea
h^{x}(\tau)&=&\frac{t}{N}\sum_{i}\left<s^{x}_{i}(\tau)\right>^{o}=\frac{t}{N}\sum_{i}\left<s^{x}_{i}\right>^{o}\nn\\
&=&\frac{t}{N}\sum_{i}\left<s^{x}_{i}\right>=tM,
\eea
does not fluctuate from site to site, and is independent of $\tau$ if we neglect subleading terms, which scale as inverse powers of $N$ .
Note that for large $N$
\bea
\sum_{i}J_{oi}^{2} \left<s^{z}_{i}(\tau)s^{z}_{i}(\tau^{\prime})\right>^{o}_{c}
\to J^2 \left[R(\tau-\tau')-q_{\rm EA}\right],
\eea
independently of the site $o$. The distribution of $h_i^z$ over the sites $i$ is the frozen field distribution,
\bea
P(y) = N^{-1}\sum_{i=1}^N \delta(y-h_i^z)
\eea
computed in the replica formalism. Thus we precisely recover the self-consistency problem for the replica diagonal, while the solution of the replica off-diagonal part furnishes the distribution $P(y)$.

For the study of the phase transition from the insulating glass phase into the superfluid, it will prove crucial to use the full low temperature solution of the SK model. However, in order to analyze properties of the mixed "superglass" phase we will restrict ourselves to a one-step approximation, which we discuss in the next section.

\subsection{Static and one-step approximation}
In order to avoid solving numerically a full self-consistent quantum problem as outlined in Eqs.~(\ref{SC}) above, we will resort to the widely used static approximation.
The latter consists in seeking a minimum of the free energy not with respect to the full function space $R(\tau)$ but, instead with respect to a constant value $R(\tau)\to R$.

A further approximation which we will use in the study of the quantum glassy phases is the one-step approximation for the structure of replica symmetry breaking. It is equivalent to assuming a step form of $q(x)$
\bea
q(x) = \Theta(x-x_1) Q_1+  \Theta(x_1-x)Q_0
\eea
and optimizing the free energy over $x_1,Q_1,Q_0$. This is expected to give qualitatively good results, especially at intermediate temperatures and close to the glass transition. Combined with the static approximation for the replica diagonal, short time part one obtains the free energy functional per spin:
\bea
\label{1RSBFE}
\beta f&=&\frac{\beta^{2}J^{2}}{4}\left[(x_1-1)Q^{2}_{1}-x_1Q^{2}_{0}+R^{2}\right] \\
&&+\frac{\beta t}{2} M^{2}-\frac{1}{x_1}\int Dy_0 \nn\\
&&\times\log\int Dy_1\left[\int Dy_R 2\cosh(\beta\sqrt{h^2_{y}+t^{2}M^{2}})\right]^{x_1}, \nn
\eea
where $h_{y}=y_0+y_1+y_R$. $Dy_{0}$, $Dy_{1}$ and $Dy_{R}$ are Gaussian measures: $Dy_0=\frac{\exp\left(-\frac{y_0^{2}}{2Q_{0}J^{2}}\right)}{\sqrt{2\pi Q_{0}J^{2}}}dy_{0}$, $Dy_1=\frac{\exp\left(-\frac{y_1^{2}}{2(Q_{1}-Q_{0})J^{2}}\right)}{\sqrt{2\pi(Q_{1}-Q_{0})J^{2}}}dy_{1}$ and $Dy_R=\frac{\exp\left(-\frac{y^2_{R}}{2(R-Q_{1})J^{2}}\right)}{\sqrt{2\pi(R-Q_{1})J^{2}}} dy_R$.

Note that $Q_1=q_{\rm EA}$ is the Edwards Anderson order parameter in the one step approximation, while $Q_{0}$ is the overlap between different spin glass states. We point out that the above free energy differs from the expression given in Ref.~\onlinecite{Gingras}, where the static approximation was not carried out correctly. This error was at the origin of several strange features of the phase diagram reported there, such as a $T$-independent transition between superfluid and superglass and a $J$-independent superfluid transition.

\subsection {1RSB free energy and self consistent equations}

Here we rewrite the one-step self-consistency equations with the help of the local field distribution.

The effective partition function of a single spin is
\bea
\label{EPF}
Z_{\rm eff}(y)&=&{\rm Tr} {\cal T}\exp\left(- S_{\rm eff}(y) \right)\\
&=&{\rm Tr} {\cal T}\exp\left( \frac{\beta^2 J^2}{2} \int_0^1\int_0^1d\tau d\tau' \right. \nn\\
&&\left.s^z(\tau) \left[R(\tau-\tau')-q_{\rm EA}\right]s^z(\tau')\right.\nn\\
&&\left.+\beta tM \int_0^1  d\tau s^x(\tau)+\beta y \int_0^1  d\tau s^z(\tau) \right). \nn
\eea

In the case of one-step replica symmetry breaking, the frozen field distribution within one pure state can be obtained by stepwise integration of the flow equations (\ref{SD1},\ref{SD2}), yielding (cf.~\onlinecite{MezardParisi})

\bea
P(y)=\int Dy_0 \frac{\int Dy_1 \delta (y-(y_0+y_1))Z^{x_{1}}_{\rm eff}(y_0+y_1)}{\int D\widetilde{y_1}Z^{x_{1}}_{\rm eff}(y_0+\widetilde{y_1})},
\eea
where $D\widetilde{y_1}$ is a Gaussian measure like $Dy_1$ with variance $(Q_{1}-Q_{0})J^{2}$.

Under the static approximation, Eq.~(\ref{EPF}) becomes
\bea
\label{effective partition function}
Z_{\rm eff}(y)=\int Dy_RZ_{\rm stat}(y+y_R),
\eea
where
\bea
Z_{\rm stat}(y)=2\cosh(\beta \sqrt{y^{2}+M^{2}t^{2}}).
\eea

One can interpret $y_R$ as a random field, which is generated by the thermal fluctuations of the non-frozen part of the magnetization.

The longitudinal and transverse magnetizations of a spin in a frozen field $y$ introduced in Eqs.~(\ref{SC}) are easily seen to be given by
\bea
\label{longitudinal magnetization}
m(y)& =&\left<s^{z}\right>_{S_{\rm eff}(y)} = \frac{1}{\beta} \frac{\partial}{\partial y}\log(Z_{\rm eff}(y)),\\
\label{transverse magnetization}
m_x(y)& =&\left<s^{x}\right>_{S_{\rm eff}(y)} = \frac{1}{\beta} \frac{\partial}{\partial (tM)}\log(Z_{\rm eff}(y)).
\eea

The saddle point equations for the Edwards-Anderson parameter $Q_{1}$ and the superfluid order parameter $M$ can now be expressed as:
\bea
\label{SC1}
Q_{1}=\frac{1}{N}\sum_{i}\left<s^{z}_{i}\right>^{2}=\int dy P(y) m^{2}(y),
\eea

\bea
\label{SC2}
M=\frac{1}{N}\sum_{i}\left<s^{x}_{i}\right>=\int dy P(y) m_{x}(y).
\eea

The saddle point equation for the parameter $R$ reads

\bea
\label{SC3}
\beta (R-Q_1) =\int P(y)\chi_{\rm loc}^{\parallel}(y) dy,
\eea
which relates the static approximation of the connected $s^z$-correlator, $R-Q_1$, to the average local susceptibility
\bea
\label{locsusc}
\chi_{\rm loc}^{\parallel}(y)=\frac{\partial m(y)}{\partial y}.
\eea

The saddle point equation for the $Q_{0}$ can be written in a similar way:
\bea
\label{SC4}
Q_{0}=\int dy_0 P(y_0;x_{1})m^{2}(y_0;x_{1}),
\eea
where
\bea
P(y_0;x_{1})&=&\frac{1}{\sqrt{2\pi Q_{0}J^{2}}} \exp(-\frac{y_0^{2}}{2Q_{0}J^{2}}),\nn\\
m(y_0;x_{1})&=&\frac{\int Dy_1Z^{x_{1}}_{\rm eff}(y_0+y_1)m(y_0+y_1)}{\int D\widetilde{y_1}Z^{x_{1}}_{\rm eff}(y_0+\widetilde{y_1})},
\eea
are discrete versions of the continuous functions $P(y,x), m(y,x)$ introduced above.

Optimizing the one-step free energy with respect to $Q_{1}$, $M$, $R$ and $Q_{0}$ yields the saddle point equations Eqs.~(\ref{SC1}-\ref{SC4}).
To capture equilibrium states, we should further extremize with respect to the Parisi parameter $x_{1}$, i.e.$\frac{\partial f}{\partial x_{1}}=0$, which yields  the further condition
\bea
&&-\frac{\beta^{2}J^{2}}{4}(Q^{2}_{1}-Q^{2}_{0})m^{2}=\int Dy_0\log\int Dy_1Z^{x_{1}}_{\rm eff}(y_0+y_1) \nn\\
&&-x_{1}\int Dy_0\frac{\int Dy_1Z^{x_{1}}_{\rm eff}(y_0+y_1)\log Z_{\rm eff}(y_0+y_1)}{\int Dy_1Z^{x_{1}}_{\rm eff}(y_0+y_1)}.
\eea

It is a useful check that upon imposing $Q_{1}=Q_{0}$, the saddle point equations for $Q_{0}$ and $Q_1$ reduce to the same replica symmetric constraint.
When $M=0$, the local field distribution, the free energy and the saddle point equations reduce to those  of the classical SK model, as it should be.

\section {Phase diagram}

Let us now study the phase diagram of our model (\ref{Hamiltonian1}). The gross features of the phase diagram we find are similar to the ones found in Refs.~\onlinecite{Gingras, Zamponi}: The low temperature phase exhibits three phases: a non-glassy superfluid at small $J/t$, an insulating (non-super-fluid) glass phase at large $J/t$, and most interestingly, a phase in between with both glassy order and superfluidity.
However, as mentioned before, we find a distinctly different behavior of the phase boundaries than Ref.~\onlinecite{Gingras}.

Moreover,  we are able to analyze the limit $T\to 0$, whose properties were inaccessible in previous works.~\cite{Gingras,Zamponi}
The latter is of particular interest in the context of the superfluid-insulator transition.

The findings of the mean field analysis are in qualitative agreement with Monte Carlo studies in finite dimensions at low but finite temperatures. The analytical approach allows for a detailed analysis of the properties of the mixed phase, and of the glass-to-superglass transition.

\begin{figure}
\includegraphics[width=3.0in]{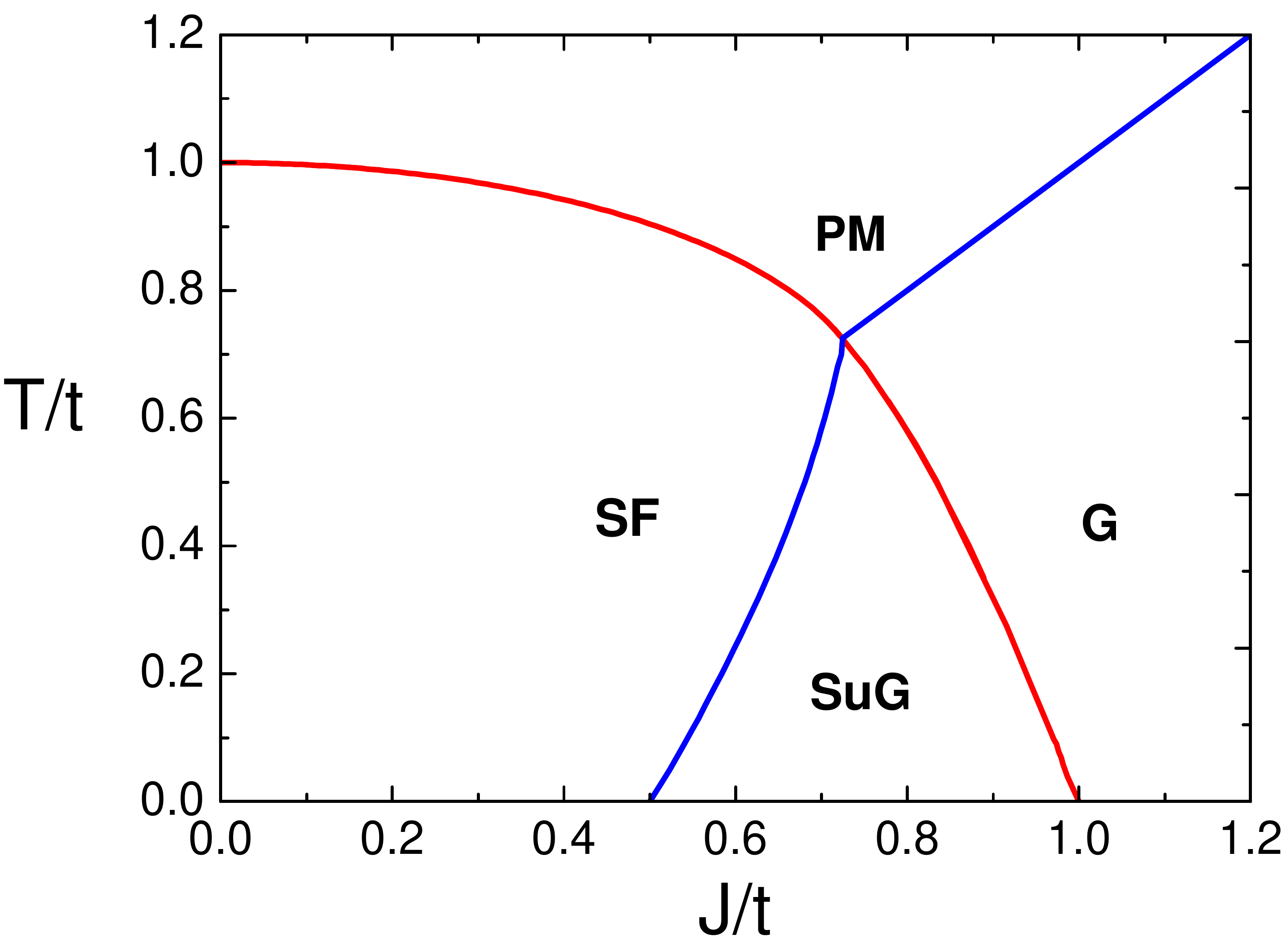}
\vspace{.2cm}
\caption{ Phase diagram of glassy hard core bosons. At high temperature, the straight blue line $T=J$ indicates the classical SK glass transition line.
The red solid line shows the superfluid phase boundary which is given by the instability condition (\ref{PMSF}). The two lines cross at the tricritical point $(T/t)_{T}=(J/t)_{T}=0.7248$. At low temperature, the blue line shows the phase boundary of the glass within the superfluid phase, as evaluated within the static approximation, cf.~(\ref{instability of glass}). The glass transition at $T=0$ occurs at $(J/t)_{g,{\rm stat}}=1/2$. The red solid line indicates the location of the onset of superfluidity within the glass phase, as evaluated within the full breaking of the replica symmetry to the instability condition (\ref{SFIC}). The superfluid transition at $T=0$ takes place at $(J/t)_s=1.00$.}
\label{f:phasediagram}
\end{figure}

\subsection{High temperature phase}
The high temperature phase is simple to describe. Since $M=0$, the system behaves identically to the paramagnetic phase of the classical SK model, and $R=1$ holds exactly. In this regime the static approximation is of course exact.

At large enough $J/t$, the leading instability upon lowering the temperature is the  classical glass transition at $T_g=J$, as mentioned earlier.
However, at small values of $J/t$, the tendency to form a superfluid wins.
The instability condition towards $XY$ symmetry breaking,
\bea
t \chi^\perp= t \frac{\partial m_{x}(y=0)}{\partial h_x}\Big|_{h_x=0} =1
\eea
can be evaluated exactly. In this expression $h_x$ is a uniform transverse field.
The transverse susceptibility is easily calculated for the replicated Hamiltonian with a Hubbard-Stratonovich transformation of the quadratic term $R\int d \tau d\tau' s^z(\tau)s^z(\tau')$. This results in the instability criterion
\bea
\label{PMSF}
\frac{1}{t} = \chi^{\perp} = \beta \frac{\int dh  e^{-h^2/2J^2} \sinh(\beta h)/\beta h}{\int dh e^{-h^2/2J^2} \cosh{\beta h}}.
\eea
The glass transition and the superfluid transition line cross at the tricritical point
\bea
T_T=J_T = t_T
\frac{\int dz  e^{-z^{2}/2} \sinh(z)/z}{\int dz e^{-z^{2}/2} \cosh{z}}
 = 0.7248\, t_T.
\eea

The result (\ref{PMSF}) does not have the familiar looking form of an average local transverse susceptibility. However, it can indeed be recast in such a way. This furnishes us a better understanding of the interaction effects in the high temperature phase, and at the same time illuminates the nature of the static approximation in the superfluid phases.

Let us rederive the above result directly from the non-replicated Hamiltonian:
\bea
H&=&-\frac{1}{2}\sum_{ij}J_{ij}s^{z}_{i}s^{z}_{j}-\frac{t}{2N}\sum_{ij}\left(s^{x}_{i}s^{x}_{j}+s^{y}_{i}s^{y}_{j}\right)\nn\\
&&\quad  -\sum_{i}h_{x}s^{x}_{i},
\eea
where $h_{x}$ is an infinitesimal field.
In a given classical Ising configuration, the spin $i$ sees an "instantaneous" local field $h^{z}_{i}=\sum_{j\neq i}J_{ij}s^{z}_{j}$, while the transverse coupling is negligible in the paramagnetic phase where $N^{-1}\sum_j\left<s_j^{x}\right>=M=0=N^{-1}\sum_j\left<s_j^{y}\right>$.
Thus the transverse susceptibility can be calculated as a site and configuration average of the susceptibility of a single spin sitting in an instantaneous field $h$,
$\chi^\perp({h})=\int^{\beta}_{0}d\tau \left<s^{x}(\tau)s^{x}(0)\right>
=\tanh(\beta h)/h$.

The thermal distribution of instantaneous local fields of the SK model has been well studied,~\cite{ThomsonSherrington} and takes the rather simple form
\bea
P_{\rm inst}(h)=\cosh(\beta h)\frac{\exp(-\frac{h^{2}}{2J^{2}}-\frac{\beta^{2}J^{2}}{2})}{\sqrt{2\pi J^{2}}}
\eea
in the paramagnetic phase. Note that the instantaneous field distribution is not a simple Gaussian, but small fields are under-represented. This phenomenon is closely related to the suppression of small fields encountered in the cavity approach to Ising systems,~\cite{replicawithoutreplicas} and is a precursor effect of the opening of the pseudogap in glassy phases at low temperatures.~\cite{Pankov, PankovDobrosavljevic, MuellerPankov}

The total transverse susceptibility is obtained as an average of the local susceptibility $\chi(h)$ over $P_{\rm inst}(h)$:
\bea
\chi^\perp &=& \int dh P_{\rm inst}(h)\chi^\perp(h) \nn\\
&=&e^{-\frac{\beta^{2}J^{2}}{2}}\int dh \frac{e^{-\frac{h^{2}}{2J^{2}}}}{\sqrt{2 \pi J^{2}}} \frac{\sinh{\beta h}}{h},
\eea
which indeed coincides with the replica result (\ref{PMSF}).

The static approximation for superfluid phases has a completely analogous effect. The approximation replaces the dynamically fluctuating exchange fields on the various sites by a random distribution of quasi static fields. The latter differs from the distribution of frozen fields (which is Gaussian at high $T$) by a random Gaussian smearing with variance $J^2(R-q_{\rm EA})$, and a reweighing factor proportional to $\cosh(\beta h)$ which accounts for the fact that a small instantaneous fields is less likely to observe on a given site, as it implies a positive  free energy fluctuation in the environment.

\subsection {Onset of glassy order within the superfluid}

The instability towards forming a glass occurs when
$J  \chi^{\parallel}=\beta J\int d\tau d\tau' R(\tau-\tau')=1$.
Within the superfluid phase it is difficult to calculate this susceptibility exactly, and we thus first resort to the static approximation, $R(\tau-\tau')\to R$. The instability of the statically approximated free energy occurs when $J \beta  R=1$, $\beta R$ being the static approximation for the longitudinal susceptibility $\chi^{\parallel} $. Within the non-glassy superfluid phase there are no frozen fields, $P(y)=\delta(y)$. Thus, from Eqs.~(\ref{SC2}-\ref{SC3}), the two relevant saddle point equations read

\bea
M=m_{x}(y=0),
\eea
\bea
\beta R=
\chi_{\rm loc}^{\parallel}(y=0),
\eea
where  $m_{x}(y=0)$ and $\chi_{\rm loc}^{\parallel}(y=0)$ are to be evaluated from Eqs.~(\ref{effective partition function}-\ref{transverse magnetization}) and (\ref{locsusc}).

They have a relatively simple low temperature limit. One verifies that it is self-consistent to assume that
\bea
\beta R =\chi^{\parallel} \to \frac{r}{t}, \quad M\to 1-m\frac{T}{t},
\eea
with finite numbers $r,m$, as $T\to 0$.

Injecting this into the above self-consistency equations, and evaluating the Gaussian integral over $y_R$ in Eq. ({\ref{effective partition function}) around the stationary point, the equations simplify to:
\bea
r &=& 1 + \frac{J^2r}{t^2-J^2r} + O(T/t),\\
m & =& \frac{J^2r}{2(t^2-r J^2)}+ O(T/t).
\eea
This yields the solution for the susceptibility $J \chi^{\parallel}=Jr/t = \frac{t}{2J}\left(1+\sqrt{1-4\left(\frac{J}{t}\right)^{2}}\right)$.

The static approximation predicts the quantum glass transition to occur at the critical point
\bea
\label{Jt_g}
\left(\frac{J}{t}\right)_{g,{\rm stat}}=\frac{1}{2},\quad\quad (T=0)
\eea
where
$J\chi^{\parallel}=1$.

It is difficult to predict whether we over- or underestimate the phase boundary with the static approximation in the superfluid phase. This is because the approximation has two competing effects with respect to the onset of glassy order.
On one hand, we approximate the dynamic longitudinal susceptibility by the static one. Since the latter is bigger, we tend to overestimate the stability of the glassy ordering of $s^z$. This effect is well-known from the SK model in a (constant) transverse field $\Gamma$.~\cite{Usadel,Huse,GrempelRozenberg} On the other hand, the static approximation underestimates quantum fluctuations of $s_x$, at least at low $T$. Indeed we see above that at $T=0$, the static approximation predicts maximal transverse order, $M=1$, independently of the value of $J/t$, while it is easy to show that quantum fluctuations around the transverse ferromagnetic state decrease the magnetization as $M=1-O((J/t)^2)$. The overestimate of $M$ leads to an underestimate of the longitudinal susceptibility, and thus of the tendency to glassy order. In view of these competing tendencies, it is hard to predict on which side with respect to Eq.~(\ref{Jt_g}) the exact glass instability will be located.

However, there is a simple way to obtain an upper bound for the quantum critical point. In the superfluid phase our model (\ref{Hamiltonian1}) is very similar to the SK model in a constant transverse field $\Gamma$,~\cite{Huse} with the difference that the effective transverse field $Mt$ is self-generated and has to be determined self-consistently. However, it is clear that the effective transverse field is always smaller than $t$. From quantum Monte Carlo results for the transverse field SK model, one knows that a quantum glass phase obtains for $J/\Gamma \geq 0.76$.~\cite{GrempelRozenberg} This implies that the model studied in the present work must certainly be in a glassy phase if $J/t \geq 0.76$. The latter value is thus an upper bound for $(J/t)_g$. Approaching from large values of $J/T$ we will find below in Eq.~(\ref{Jts}) that the non-superfluid glass phase becomes unstable towards superfluidity already at $(J/t)_s=1.00$. Hence, we conclude that a phase with both superfluid and glassy order parameters exists for a substantial range of parameters covering at least the interval  $0.76\leq J/t \leq 1.00$.

\subsection {Superfluid instability within the insulating glass phase}
\subsubsection{Instability criterion}
Our discussion of the phase boundaries will be complete, once we have addressed the superfluid instability with in the glass phase at large $J/t$.
The instability condition reads
\bea
t \int dy P(y)\frac{\partial m_x(y)}{\partial h_x}\Big|_{h_x=0}=1,
\eea
where $P(y)$ is the non-trivial distribution of frozen local fields in the classical glass phase of the SK model.
The properties of $P(y)$ are well studied, and turn out to be crucial to understand the low temperature behavior of the phase boundary and the physics of the glassy superfluid-to-insulator quantum phase transition.

We recall that in the non-superfluid glass phase the static approximation is exact with $R=1$, so that the instability criterion can be expressed in the form,

\bea
\label{SFIC}
t\int dy P(y)\frac{\int Dy_R \sinh(\beta (y+y_R)) \frac{1}{y+y_R}}{\int Dy_R \cosh(\beta (y+y_R))}=1,
\eea
where $Dy_R=\frac{1}{\sqrt{2\pi(1-q_{\rm EA})J^{2}}}\exp(-\frac{y^2_{R}}{2(1-q_{\rm EA})J^{2}})$. This condition can be expressed in terms of the instantaneous field distribution as
\bea
\label{BCS}
t\int dh P_{\rm inst}(h) \frac{\tanh(\beta h)}{h} = 1,
\eea
where
\bea
P_{\rm inst}(h) = \int P(y) dy\frac{\cosh{\beta h}}{\cosh{\beta y}}\frac{\exp\left(-\frac{\beta(h-y)^{2}}{2h_{O}}-\frac{\beta h_{O}}{2}\right)}{\sqrt{2\pi h_{O}/\beta}}.
\eea
is the instantaneous field distribution, which was first derived in Ref.~\onlinecite{ThomsonSherrington}.
The term $h_{O}=\beta J^{2}(1-q_{\rm EA})$ is known as Onsager's back reaction. Eq.~(\ref{BCS}) can be recognized as a BCS-equation,
where the instantaneous field distribution $P_{\rm inst}(h)$ takes the role of the density of states.

The temperature dependent local field distribution can be obtained from a numerical solution of the self-consistent set of full RSB equations (\ref{SD1}-\ref{SCE}), from which the phase boundary of the insulator-to-superfluid transition is deduced . This yields the  solid [red] line in Fig.~\ref{f:phasediagram}). For comparison we also evaluate the phase boundary within a one-step approximation, which works well at moderate temperatures. However, it fails badly at low $T$ where a non-physical reentrance of the superfluid instability would be predicted, and the quantum phase transition at $T\to 0$ is completely missed.

We note in passing that the thermodynamics of the insulating phase is essentially classical because of the scaling of the transverse coupling as $t/N$. If instead $t$ were random and scaled as $1/\sqrt{N}$, the glass phase would also exhibit quantum fluctuations and would not reduce to the purely classical SK model. In that case, the analysis of the transition would become much more complicated.
However, even though the thermodynamics can be obtained by a purely classical saddle point computation, one should not conclude that excitations do not have any quantum dynamics.

\subsubsection{Low temperatures and quantum phase transition}

At low temperatures, the most prominent feature of the local field distribution $P(y)$ is a linear pseudogap which opens at small fields. The latter is required to assure the stability of the glass phase,~\cite{ThoulessAnderson77,PalmerPond} in a very similar manner as the Efros-Shklovskii Coulomb gap arises in electron glasses with unscreened, long range $1/r$ interactions.~\cite{MuellerIoffe, MuellerPankov} More precisely, it is known that $P(y)= \alpha |y| +O(T) $ with $\alpha = 0.301$ for fields in the range $T\ll |y| \ll J$, while the distribution decays like a Gaussian for $|y|\gg J$. At zero temperature the pseudogap extends down to $y=0$ (i.e., the chemical potential in the terminology of hardcore bosons), while at finite but low temperatures $T\ll J$, $P(y)$ assumes a scaling form
$P(y) = Tp(y/T)$ with $P(0) = {\rm const.}$ and $p(x \gg 1)=\alpha |x|+ {\rm const.}$.~\cite{MuellerPankov, Pankov}

This scaling form asserts that only a fraction of $~(T/J)^2$ is thermally active. Therefore the Edwards-Anderson parameter tends to $1$ as $1-q_{\rm EA}\sim (T/J)^{2}$. Accordingly, as $T\to 0$ there is no difference between the distribution of frozen and instantaneous fields,
$P_{\rm inst}(h)$, since no thermal fluctuations are left. In this limit the instability condition (\ref{SFIC}) for onset of superfluidity then takes the form:
\begin{equation}
\label{Jts}
t_{s}\int dy \frac{P(y; T=0)}{|y|}=1.
\end{equation}

Using the above mentioned features of $P(y)$ at low $T$ one can easily obtain a rough estimate for the superfluid-insulator transition point as $(J/t)_s\simeq 1.05\pm 0.1$. However, since the precise value is also sensitive to the part of $P(y)$ at high fields, $y\geq J$, a full numerical  evaluation of the condition (\ref{Jts}) is necessary to obtain the exact location of the quantum critical point. Using high precision data for $P(y;T)$ at low $T$ from Ref.~\onlinecite{Leuzzi}, we find $(J/t)_{s} \simeq 1.00\pm 0.01$.

We emphasize an important difference between the quantum phase transition we have found here and a standard BCS transition. The latter, in the presence of a constant low energy density of states always yields a finite $T_c$, even though it becomes exponentially small in $1/t$ for small $t$. In our glassy system the situation is fundamentally different in that the frustrated interactions suppress the density of states around the chemical potential with $P(y\to 0)\to 0$. This quenches the tendency for superfluidity and allows for a superfluid-to-insulator transition at a finite value of $t$, even in the mean field limit of $N\to \infty$ which we consider here.

This has important consequences for the nature of excitations and transport properties across the superfluid-insulator transition. In particular, the transition to the Bose insulator is accompanied by the Anderson localization of lowest energy excitations, whereas higher energy excitations remain delocalized relatively far into the insulator.~\cite{fn1}
We believe that the physics revealed by this mean field model is relevant for Coulomb frustrated bosonic systems which undergo a transition from a superfluid to a Bose glass state in finite dimensions. This will be discussed in detail elsewhere.~\cite{XMinprep}

It is interesting to compare our mean field predictions for the phase diagram with the $3\textrm{D}$ quantum Monte Carlo (QMC) simulation results reported in  Ref.~\onlinecite{Gingras}. The mean field predictions for the quantum critical points actually match the numerical results surprisingly well.
The latter were done for the Hamiltonian
\bea
H&=&-\sum_{\left<i,j\right>}V'_{ij}(n_{i}-1/2)(n_{j}-1/2)\nn\\
&&-t'\sum_{\left<i,j\right>}\left(b^{\dag}_{i}b_{j}+\textrm{h}.\textrm{c}.\right),
\eea
with binary disorder, $V'_{ij} =\pm V'$ with equal probability. Contact with the mean field model (\ref{boseHam}) is made by replacing the coordination number with $N\to z=6$ for the $3\textrm{D}$ cubic lattice, and taking a Gaussian disorder with the same variance, $V^2/z=V'^2$, as well as a hopping $t_b/z=t'$.

Recalling the dictionary (\ref{dict}), the mean field estimate of the superglass to glassy insulator quantum phase transition is
\bea
\left(\frac{V'}{t'}\right)^{\textrm{MF}}_{s}&=&\left(\frac{V/\sqrt{z}}{t_{b}/z}\right)_{s}=\left(\frac{4J/\sqrt{z}}{2t/z}\right)_{s} \\
 &=&2 \sqrt{z}\left(\frac{J}{t}\right)_{s}\simeq 4.9 \approx \left(\frac{V'}{t'}\right)^{\textrm{QMC}}_{s}\backsimeq 5, \nn
\eea
which comes close to the extrapolation of QMC results to $T=0$.
The transition point between superglass and non-glassy superfluid is estimated from the static approximation as
\bea
\left(\frac{V'}{t'}\right)^{\textrm{MF}}_{g}=2 \sqrt{z}\left(\frac{J}{t}\right)_{g,{\rm stat}}\simeq 2.45 \nn\\
\approx\left(\frac{V'}{t'}\right)^{\textrm{QMC}}_{g}\backsimeq 3.2.
\eea
This indicates that the static approximation overestimates the stability of the superglass phase, similarly as what is known from the mean field version of the transverse field Ising spin glass.

The mean field prediction (with static approximation) for the interaction-to-hopping ratio $(V'/t')_{T}$ at the tricritical point is rather good, too,
\bea
\left(\frac{V'}{t'}\right)^{\textrm{MF}}_{T}=2 \sqrt{z}\left(\frac{J}{t}\right)_{T,{\rm stat}}\simeq 3.55 \nn\\
\approx\left(\frac{V'}{t'}\right)^{\textrm{QMC}}_{T}\backsimeq 3.8.
\eea
While the tricritical ordering temperature is overestimated by a factor of 2 (similarly as in the classical Ising spin glass)~\cite{BinderYoungReview}
\bea
\left(\frac{T}{t'}\right)^{\textrm{MF}}_{T}
= \frac{z}{2}\left(\frac{T}{t}\right)_{T}\simeq 2.2, \\
\left(\frac{T}{t'}\right)^{\textrm{QMC}}_{T}\backsimeq 1.1.
\eea

\subsection{Robustness of the phase diagram to random field disorder}

In the previous sections we have seen that the model (\ref{Hamiltonian1}) possesses an intermediate phase which is simultaneously superfluid and glassy. We have determined the phase boundaries as instability lines, assuming second order phase transitions. Indeed it seems unlikely that any of the instabilities could be preempted by a first order transition. Since the superfluid to insulator transition at $(J/t)_s$ is of particular interest, we provide further arguments in this section that the parts of the phase diagram related to the phase boundary of the non-superfluid glass remain robust when disorder potentials, i.e. random fields $\epsilon_i$ of variance $W^2$, are restituted to the model. In particular we will show that glass and superfluid transition lines meet at a tricritical point at finite temperature $T_T/J$ and $(J/t)_T$. Further we determine the superfluid instability of the glass phase at $T=0$ and show that it always occurs at a larger ratio $(J/t)$ than the tricritical point, $(J/t)_s> (J/t)_T$. This suggests that for any $W$ the transition line between non-superfluid and superfluid glass is not reentrant as a function of temperature. The absence of reentrance in turn suggests that the quantum phase transition out of the insulating glass remains second order, independent of the strength of the disorder potential.

The Hamiltonian with a disorder potentials reads
\bea
\label{Hamiltonian2}
H=-\sum_{i<j}J_{ij}s^{z}_{i}s^{z}_{j}-\frac{t}{N}\sum_{i<j}\left(s^{x}_{i}s^{x}_{j}+s^{y}_{i}s^{y}_{j}\right)+\sum_{i}\epsilon_{i}s^{z}_{i}.
\eea
The disorder potential breaks the $Z_{2}$ symmetry, therefore $Q_{a\neq b}\neq 0$ already in the high temperature phase, where it assumes a constant replica symmetric value $Q_0$.
The glass phase occurs at the Almeida-Thouless instability, which is given by: ~\cite{MuellerPankov}
\bea
\label{tricritical temperature}
\beta^{2}J^{2} \int dy \frac{P_{W}(y)}{\cosh^{4}(\beta y)} =1,
\eea
where
\bea
P_{W}(y)=\frac{\exp\left(-\frac{y^{2}}{2(W^{2}+J^{2}Q_{0})}\right)}{\sqrt{2\pi(W^{2}+J^{2}Q_{0})}},
\eea
and $Q_0$ satisfies the self-consistent equation
\bea
Q_{0}=\int P_{W}(y)\tanh^{2}(\beta y)dy.
\eea

The instability towards the superfluid phase is instead determined by
\bea
\label{tricritical hopping}
1/t=\int dh P_{\rm inst}(h)\frac{\tanh(\beta h)}{h},
\eea
where
\bea
\label{insloc}
P_{\rm inst}(h)&=&\int dy P_{W}(y)\frac{\cosh{\beta h}}{\cosh{\beta y}}\nn\\
&&\times\frac{\exp\left(-\frac{\beta (h-y)^{2}}{2h_{O}}-\frac{\beta h_{O}}{2}\right)}{\sqrt{2\pi h_{O}/\beta }},
\eea
with the Onsager field $h_{O}=\beta J^{2}(1-Q_{0})$.~\cite{ThomsonSherrington,MuellerPankov}

The glass and superfluid transition lines meet at a tricritical point at $(T/J)_T$ and $(J/t)_T$ which are to be evaluated from Eqs.~(\ref{tricritical temperature}-\ref{insloc}).

In the limit $W/J\gg 1$, one finds the tricritical temperature $(T/J)_T=\frac{4}{3\sqrt{2\pi}} \frac{J}{W}+O(\frac{J^{2}}{W^{2}})$ and $\beta_{T}h_{O}\rightarrow3/2$, as $W/J\rightarrow \infty$.

The superfluid transition at $T=0$ is given by the condition
\bea
\label{critical hopping}
(1/t)_{s}=\int dh \frac{P_{\rm inst}(h;T=0)}{|h|}.
\eea
In the limit $W/J\gg 1$, $P_{\rm inst}(h;T=0)$ is known to have a simple structure :
\bea
\label{instfield at zeroT}
P_{\rm inst}(h;T=0)=\left \lbrace \begin{array}{cc} \alpha |h|/J^{2}, & |h|\ll h^{\star}, \\ \frac{\exp\left(\frac{-(h-\gamma J^{2}/W)^{2}}{2 W^{2}}\right) }{\sqrt{2\pi W^{2}}}, & |h|\gg h^{\star}, \end{array}\right.
\eea
with a smooth crossover between the two limiting forms around $\frac{h^{\star}}{J}=\frac{1}{\alpha \sqrt{2\pi}}\frac{J}{W}+O(\frac{J^{2}}{W^{2}})$.
The value of the constant $\gamma=O(1)$ can be estimated by the normalization condition $\int dh P_{\rm inst}(h;T=0)=1$, but will be irrelevant below.

For $W/J\gg1$, $(J/t)_{s}$ and $(J/t)_{T}$ both behave as $\frac{4}{\sqrt{2\pi}}\frac{\log \left(W/J\right)}{W/J}$ to leading order. Their difference scales like $c J/W$. The coefficient $c$ can be evaluated easily by  rescaling  the variables $\beta_{T}h=\hat{h}$ and $\beta_{T}y=\hat{y}$,
\bea
\label{subleading}
c&=&\lim_{W/J\rightarrow \infty}\frac{W}{J}\left[(J/t)_{s}-(J/t)_{T}\right]\nn\\
&=&\sqrt{\frac{2}{3}}\int \frac{d\hat{h}}{\hat{h}}\left[\hat p(\hat{h};T=0)-f(\hat{h})\right],
\eea
where
\bea
\label{p(h,T=0)}
\hat p(\hat{h};T=0) &=&\lim_{W/J\to \infty} \sqrt{\frac{3}{2}}W P_{\rm inst}(\hat{h}/\beta_T;T=0)
\eea
and
\bea
f(\hat{h})=\int\frac{d\hat{y}}{\sqrt{2\pi}}\frac{\sinh(\hat{h})}{\cosh(\hat{y})}\frac{\exp\left(\frac{-(\hat{h}-\hat{y})^{2}}{3}-\frac{3}{4}\right)}{\sqrt{2\pi}},
\eea
and we have used $\beta_Th_O\to 3/2$. We approximate Eq.~(\ref{instfield at zeroT}) by extending the formula all the way to $h^{\star}$ and neglecting the shift of field $h$ and we get :
\bea
\label{estimationofp(h,T=0)}
\hat p(\hat{h};T=0)\approx \left \lbrace \begin{array}{cc} \frac{2\alpha \hat{h}}{\sqrt{3\pi}}, & |\hat{h}|\leq \frac{3}{4\alpha}, \\ \sqrt{\frac{3}{4\pi}}, & |\hat{h}|\geq \frac{3}{4\alpha}, \end{array}\right.
\eea
Evaluating Eq.~(\ref{subleading}) numerically, using the estimate Eq.~(\ref{estimationofp(h,T=0)}) , one obtains $c=0.231>0$, establishing that $(J/t)_{s}>(J/t)_{T}$ even in the presence of strong disorder. We point out that Eq.~(\ref{estimationofp(h,T=0)}) overestimates Eq.~(\ref{p(h,T=0)}), but this overestimation should be much smaller than $c=0.231$.

\begin{figure}[h]
\includegraphics[width=3.0in]{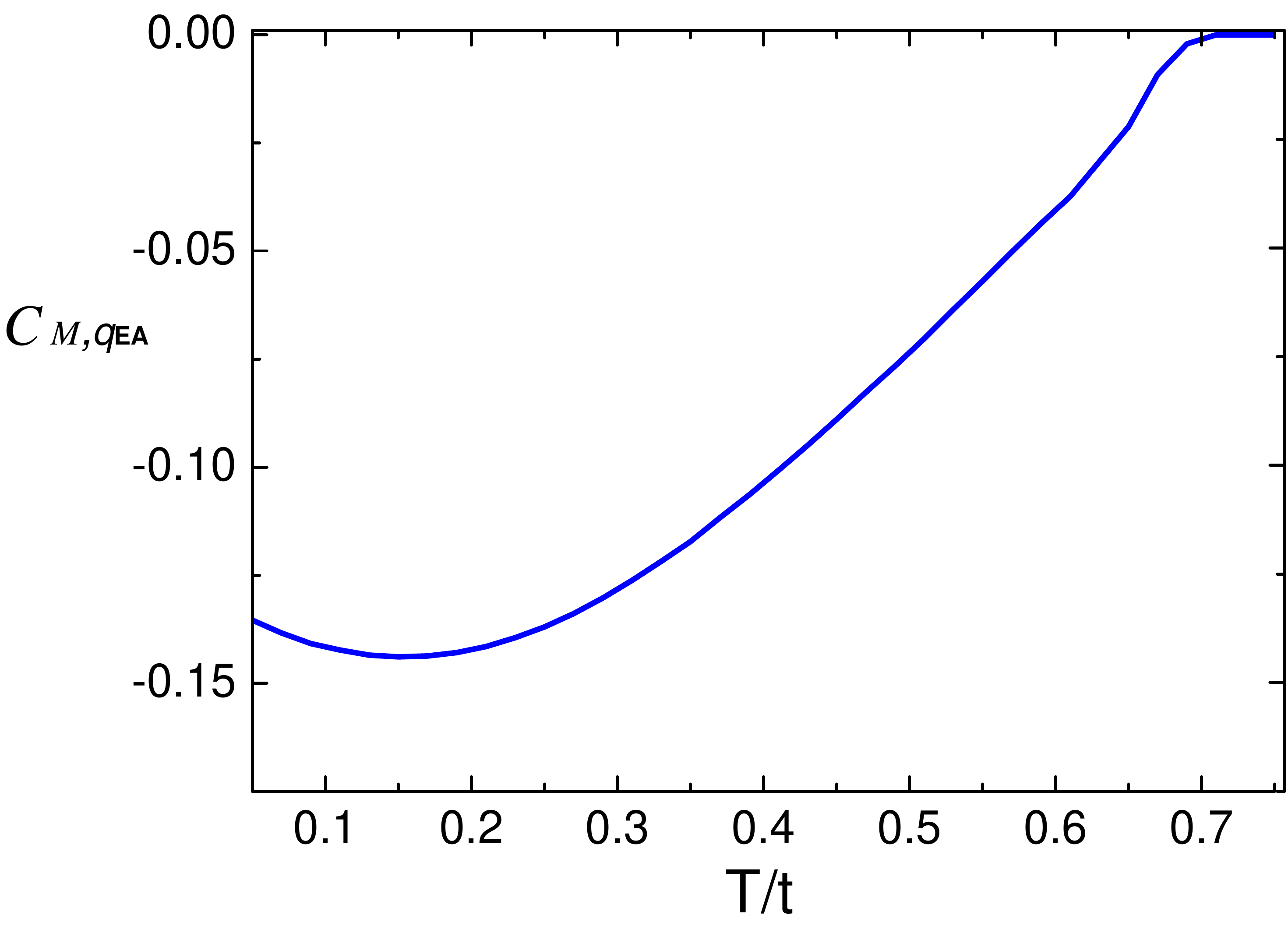}
\vspace{.2cm}
\caption{Cross-correlation between the local oder parameters for superfluidity and glassy order, respectively. The correlations are evaluated from Eq.~(\ref{orderparametercorrelation}) in the temperature range $0.05<T/t<0.75$ at fixed disorder $J/t=0.724$, close to the ratio corresponding to the tricritical point. The local order parameters are anticorrelated, the maximal anticorrelation occurs at intermediate temperatures. }
\label{f:correlation}
\end{figure}

\section{Properties of the super-glass phase}
Having established the phase diagram of the model, we now focus on the properties of the bulk of the "superglass" phase. There the interplay between temperature, glassy order and superfluid order induce several interesting phenomena which potentially survive also in finite dimensional models of frustrated bosons.
In the following, we investigate how the glassy and superfluid orders evolve with temperature, and how they are locally correlated.
\subsection{Competition between glassy and superfluid order}
\label{nonmonotonicity}
While the effective transverse field $h_i^{x}=Mt$ is uniform for every site, the frozen longitudinal field, $h^{z}_{i}$ depends on the site (and on the pure state in which the system is frozen). Therefore, the magnetization of the local spin $s_i$ due to the local field $\overrightarrow{h_{i}}=(h_i^{x},h^{z}_{i})$ fluctuates from site to site. It is interesting to study  the correlation of the local magnetization, whose components are the local order parameters of the glassy and the superfluid order, respectively. More precisely, we investigate the following correlation function:
\bea
\label{orderparametercorrelation}
C_{M,q_{\rm EA}}&\equiv&\frac{\frac{1}{N}\sum_{i}\left<s^{x}_{i}\right>\left<s^{z}_{i}\right>^{2}-(\frac{1}{N}\sum_{i}\left<s^{x}_{i}\right>)(\frac{1}{N}\sum_{i}\left<s^{z}_{i}\right>^{2})}{(\frac{1}{N}\sum_{i}\left<s^{x}_{i}\right>)(\frac{1}{N}\sum_{i}\left<s^{z}_{i}\right>^{2})}\nn\\
&=&\frac{\int dy P(y)m_{x}(y)m^{2}(y)-Mq_{\rm EA}}{Mq_{\rm EA}}.
\eea
We have evaluated the correlation function (\ref{orderparametercorrelation}) within the static 1-step RSB approximation in the center of the superglass phase ($J/t=0.724$) as a function of temperature ($0.05<T/t<0.75$), see Fig.~\ref{f:correlation}. Not surprisingly, the correlation is negative, since glassy and superfluid orders compete with each other.
Indeed, one easily checks that for every pair of sites $(i,j)$ it holds that if $\left<s^{z}_{i}\right>^{2} < \left<s^{z}_{j}\right>^{2}$ then $\left<s^{x}_{i}\right> > \left<s^{x}_{j}\right>$.  The maximal amplitude of the normalized correlation $C_{M,q_{\rm EA}}$ is only of order $\approx 0.1$, suggesting that in the superfluid phase the non-uniformity of the two local order parameter fields is actually not very strong. It may be that the 1-step approximation underestimates these correlations a bit.
The relative weakness of the anticorrelations might be the reason why they have not been noticed in the quantum Monte Carlo studies of Refs.~\onlinecite{Gingras, Zamponi}.

\subsection{Non-monotonicity of the superfluid order}
In the superglass phase, the glass order parameter $q_{\rm EA}= Q_{1}$ monotonously decreases with increasing temperature, as one should expect. However, surprisingly, the superfluid order parameter $M$ exhibits non-monotonic behavior with a maximum at an intermediate crossover temperature $T_m$, as shown in Fig.~\ref{f:orderparameter}. Below $T_{m}$, the superfluid order parameter $M$ decreases, anomalously, when lowering the temperature. Above $T_{m}$, $M$  decreases with increasing temperature as usual in a standard superfluid.

This phenomenon is related to the anti-correlation between glassy and superfluid order discussed in the previous section. While on one hand, thermal fluctuations tend to diminish both glassy and superfluid order, there appears to be a low temperature regime $T<T_{m}$, where quantum fluctuations of the superfluid order are dominant. Due to the competition between the glassy and the superfluid order, the thermally induced decrease of the glassy order enhances the superfluid order. This effect dominates over the direct thermal effects on the superfluidity.

It seems natural that it is the superfluid order which undergoes such non-monotonic behavior, rather than the glassy order. Indeed, we expect the latter to react  less sensitively to the diminution of quantum fluctuations due to decreasing transverse fields

We note that also the local order parameter correlations $C_{M,q_{\rm EA}}$ exhibit a non-monotonous behavior within our static 1-step approximation, as shown in Fig.~\ref{f:correlation}. The absolute value of $C_{M,q_{\rm EA}}$ increases with temperature at very low temperatures, and decreases at higher temperatures. This can be seen again as a consequence of the non-monotonicity of the superfluid order. At fixed $T$, the larger $h^{x}$ the stronger the normalized anticorrelation $C_{M,q_{\rm EA}}$. Since $h^{x}=tM$ initially increases with $T$, it is natural to expect an increasing $C_{M,q_{\rm EA}}$ until eventually thermal fluctuations become dominant and diminish $C_{M,q_{\rm EA}}$.

\begin{figure}[h]
\includegraphics[width=2.8in]{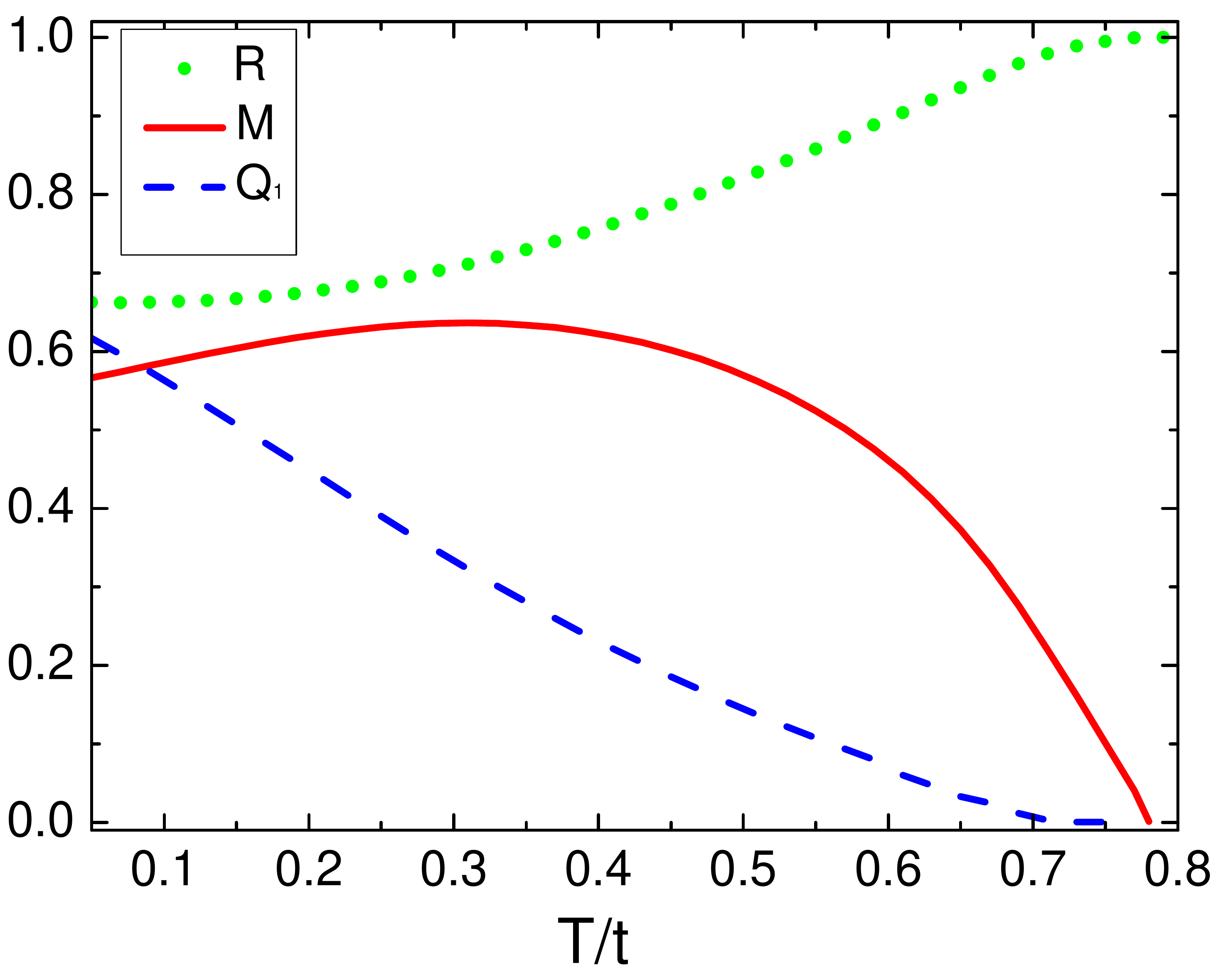}
\vspace{.2cm}
\caption{ (color online) The order parameters in the superglass phase as a function of temperature $0.05<T/t<0.79$ at the fixed disorder $J/t=0.724$.  The blue dashed line indicates the Edwards-Anderson order parameter $Q_{1}$ (in 1-step approximation), which monotonously decreases with increasing temperature. The superfluid order parameter $M$ (red solid line) exhibits non-monotonic behavior. The long time (static) on-site charge correlation $R$, (green dotted line) becomes $1$ in the disordered high $T$ phase. }
\label{f:orderparameter}
\end{figure}

\section{Discussion}

In this paper we have analyzed a fully connected mean field model. The full connectivity is not a real limitation, however. Indeed, one can generalize the model to a highly connected Cayley tree. While this does not affect the thermodynamics of the model, this generalization allows for the study of localization and delocalization of excitations, since this model is endowed with a notion of distance. The analysis of localization properties is of particular interest in the vicinity of the superglass-to-insulating glass quantum phase transition, where the boson system collectively delocalizes into a superfluid at low energies.
The nature of higher energy excitations in the insulator are crucially affected by the suppression of low energy states in the glass, leading to a non-trivial  excitation spectrum at the glassy SI transition. The details of this analysis will be reported elsewhere.~\cite{XMinprep}

What features of the mean field model should be expected to carry over to finite dimensions? In the present model we find a genuine insulating phase at $T=0$, which suppresses the superfluidity, due to the strong self-generated on-site disorder within the glassy phase. A crucial ingredient for the suppression of superfluidity is the linear pseudogap within the glass phase. A very similar pseudogap  is known to occur in disordered Coulomb interacting systems, where it is due to unscreened $1/r$ interactions between charged particles. This Coulomb gap may well be of importance in strongly disordered superconductors and play a significant role in the competition between glassy insulating behavior and superconductivity. In particular, in materials with strong negative $U$ centers, one may think of preformed electron pairs constituting hard core bosons which interact with Coulomb repulsions.~\cite{Twocomponent} The power law suppression of the low energy density of states makes it likely that the superfluid condensate is entirely destroyed once the hopping becomes too small. For short range interactions the density of states is merely reduced at low energy, but does not tend to zero. On a Cayley tree of very large connectivity this will always lead to delocalization, unless the hopping $t$ is scaled down logarithmically with the connectivity. In finite dimensions, however, sufficiently strong disorder is known to suppress superfluidity,~\cite{MaLee} and thus one may expect that  at sufficiently large ratios $J/t$, the disordered boson model  will localize due to {\em spontaneously} created, frozen-in local fields. Such a conclusion may be suggestive from a straight extrapolation of the quantum Monte Carlo results of Ref.~\onlinecite{Gingras} to $T=0$, but it seems difficult to exclude a scenario in which $T_c$ becomes merely exponentially small with $J/t$. A more careful analysis will be necessary to settle this question in finite dimensional, short range interacting glasses.

As for the coexistence phase, the "superglass", the numerical data~\cite{Gingras,Zamponi} provides evidence that it exists also in finite dimensions. It would be interesting to confirm and quantify the local anticorrelation of order parameters in such simulations. From our mean field analysis one expects that the anticorrelation is in fact relatively weak. A further non-trivial prediction with measurable consequences is the non-monotonicity of the superfluid order parameter, which should translate into an equivalent non-monotonicity of the superfluid stiffness as a function of temperature. This non-monotonicity has its origin in the softening of the glassy order  at low $T$, a feature which may potentially survive in finite dimensions, especially when the lattice connectivity is large, or the interactions are not too short ranged. We should caution though that we obtained this effect by employing a static approximation and a replica symmetry breaking at the one-step level only. However, we believe that it is a real feature of the model.

As discussed earlier, a number of experiments have already shown promising indications of possible coexistence of glassy order with superfluidity.
We hope that our analysis will help to unambiguously identify such phases in experiments. Note that finding an experimental system exhibiting a glassy superfluid-insulator transition might also be of great interest to study the intricate interplay of interactions and disorder with respect to glassy ergodicity breaking, and quantum ergodicity breaking, i.e. Anderson localization.

\section{Acknowledgment}
We thank L. Leuzzi for providing us the high precision data on the local field distribution $P(y)$ of the SK model at finite temperature.
We thank L. Foini, M. Gingras, F. Zamponi, A. W. Sandvik for useful discussions.

\vspace{8.0cm}

\appendix
\section*{APPENDIX: Derivation of RS free energy}

\label{app:RSfreeenergy}
With standard replica trick~\cite{MezardParisi} we can get the RS free energy:
\bea
\beta f&=&-\frac{J^{2}\beta^{2}}{4}Q^{2}+\frac{t\beta}{2}M^{2}+\frac{J^{2}\beta^{2}}{4}\int^{1}_{0}\int^{1}_{0}d\tau d\tau^{\prime}R^{2}(\tau,\tau^{\prime})\nn\\
&&-\lim_{n\rightarrow 0}\frac{1}{n}\log {\rm Tr} {\cal T}\exp \left[\frac{J^{2}\beta^{2}}{2}Q\left(\sum_{a}\int^{1}_{0}dts^{z}_{a}(\tau)\right)^{2} \right.\nn\\
&&\left.+\frac{J^{2}\beta^{2}}{2}\sum_{a}\int^{1}_{0}d\tau\int^{1}_{0}d\tau^{\prime}R(\tau,\tau^{\prime})s^{z}_{a}(\tau)s^{z}_{a}(\tau^{\prime})\right.\nn\\
&&\left.-\frac{J^{2}\beta^{2}}{2}Q\sum_{a}\left(\int^{1}_{0}dts^{z}_{a}(\tau)\right)^{2}\right. \nn\\
&&\left.+t\beta\sum_{a}\int^{1}_{0}d\tau M^{x}_{a}s^{x}_{a}(\tau)\right].
\eea
Under static approximation: $R(\tau,\tau^{\prime})=R$, we have
\bea
\beta f&=&-\frac{J^{2}\beta^{2}}{4}Q^{2}+\frac{t\beta}{2}M^{2}+\frac{J^{2}\beta^{2}}{4}R^{2}\\
&&-\lim_{n\rightarrow 0}\frac{1}{n}\log {\rm Tr} {\cal T} \exp\left[\frac{J^{2}\beta^{2}}{2}Q\left(\sum_{a}\int^{1}_{0}dts^{z}_{a}(\tau)\right)^{2} \right. \nn\\
&&\left.+\frac{J^{2}\beta^{2}}{2}(R-Q)\sum_{a}\left(\int^{1}_{0}dts^{z}_{a}(\tau)\right)^{2}\right.\nn\\
&&\left.+t\beta\sum_{a}\int^{1}_{0}dtM^{x}_{a}s^{x}_{a}(\tau)\right].\nn
\eea
According to Hubbard-Stratonovich transformations, we linearize the quadratic terms $\left(\sum_{a}\int^{1}_{0}dts^{z}_{a}(\tau)\right)^{2}$ and $\left(\int^{1}_{0}dts^{z}_{a}(\tau)\right)^{2}$ by introducing extra fields $y_{0}$ and $y_{R}$:
\bea
\beta f&=&-\frac{J^{2}\beta^{2}}{4}Q^{2}+\frac{J^{2}\beta^{2}}{4}R^{2}+\frac{t\beta}{2}M^{2} \\
&&-\int Dy_0\log \int Dy_R{\rm Tr}\exp\left[\beta(y_0 s^{z}+y_R s^{z}+t M s^{x})\right] \nn\\
&&=-\frac{J^{2}\beta^{2}}{4}Q^{2}+\frac{J^{2}\beta^{2}}{4}R^{2}+\frac{t\beta}{2}M^{2}\nn\\
&&-\int Dy_0\log \int Dy_R \cosh\left(\beta\sqrt{\left(y_0+y_R \right)^{2}+t^{2}M^{2}}\right).\nn
\eea
One can get (\ref{1RSBFE}) following the similar steps above.


\begin{thebibliography}{99}
\bibitem{fn1}
{The notion of delocalization requires a straightforward extension of the present model to a lattice model,
which is not fully connected but maintains a large coordination number $N$.}
\end{thebibliography}
\end{document}